\documentclass[preprint,superscriptaddress,aps,showpacs]{revtex4}
\usepackage{amsfonts}
\usepackage{graphics}
\usepackage{graphicx}
\usepackage{slashed}	
\newcommand{\be}{\begin{equation}}
\newcommand{\ee}{\end{equation}}
\newcommand{\en}{\end{equation}}
\newcommand{\ba}{\begin{eqnarray}}
\newcommand{\ea}{\end{eqnarray}}
\newcommand{\bea}{\begin{eqnarray}}
\newcommand{\eea}{\end{eqnarray}}
\newcommand{\pa}{\partial}

\def\As{A\!\!\!/}
\def\ks{k\!\!\!/}

\def\ps{p\!\!\!/}

\def\us{u\!\!\!/}
\def\vs{v\!\!\!/}

\def\ds{\partial\!\!\!/}

\begin{document}

\title{On one-loop corrections in the non-minimal dimension-five extension of QED}
\author{T. Mariz}
\affiliation{Instituto de F\'\i sica, Universidade Federal de Alagoas, 57072-270, Macei\'o, Alagoas, Brazil}
\email{tmariz@fis.ufal.br}

\author{J. R. Nascimento}
\affiliation{Departamento de F\'{\i}sica, Universidade Federal da Para\'{\i}ba\\
 Caixa Postal 5008, 58051-970, Jo\~ao Pessoa, Para\'{\i}ba, Brazil}
\email{jroberto, petrov@fisica.ufpb.br}

\author{A. Yu. Petrov}
\affiliation{Departamento de F\'{\i}sica, Universidade Federal da Para\'{\i}ba\\
 Caixa Postal 5008, 58051-970, Jo\~ao Pessoa, Para\'{\i}ba, Brazil}
\email{jroberto, petrov@fisica.ufpb.br}
\author{H. Belich}
\affiliation{Departamento de F\'{\i}sica e Qu\'{\i}mica, Universidade Federal do Esp\'{\i}%
rito Santo, Av. Fernando Ferrari, 514, Goiabeiras, 29060-900, Vit\'{o}ria,
ES, Brazil.}
\email{belichjr@gmail.com}

\begin{abstract}
In this paper, we describe the generation of the CPT-even, aether-like terms
via the new CPT-even magnetic-like coupling. We carry out a study the
loop corrections generated by this coupling. Previous investigations has
been initiated on this issue and we have extended them to studying
of higher-point functions, of quantum corrections to
vertices of the interaction and to two-point function of the spinor field.
\end{abstract}


\maketitle

\section{Introduction}

The discovery of the Higgs boson in 2013 confirmed the physical
interpretation of particles as the manifestation of the
excitations of quantized fields described by the Standard Model of the
fundamental interactions. Despite the successful program of the Standard
Model, the proposal depends on 19 parameters that need to agree with
experiments in order to indicate if it is really an appropriate theory.
Moreover, the unification of the fields in the Standard Model remains
incomplete. The electromagnetic and weak fields are unified in the
Weinberg-Salam model \cite{novaes}, the strong field is apart from them (the
gauge sector of the Standard Model is $SU(3)\times SU(2)\times U(1)$), and
the gravitational interaction is conserved apart. Besides, in the Standard
Model we have the problem of the mass of the Higgs boson that diverges
heading towards the Planck scale, and we do not have a satisfactory response
to the problem of matter-antimatter asymmetry. Neutrinos appear to be
massless particles in the Standard Model, which contradicts experiments of
the nineties that observed neutrinos oscillation.

These restricted possibilities of SM extensions motivated
proposal to create effective models which can give us hints about a more
fundamental theory. In 1989, Kosteleck\'{y}
and Samuel \cite{sam} demonstrated that the spontaneous Lorentz symmetry breaking emerges in the low-energy limit of string theory. In this case some vector and/or tensor fields acquire non-zero vacuum expectation values, and, as a result, the Lorentz
symmetry turns out to be spontaneously broken \cite{ens}.

The idea of this proposal is to investigate the presence of background
fields which causes anisotropy in space-time, since it can affect the
physical properties of the particles. The detection of fields that comes
from this breaking of symmetry can be viewed through a change in the kinetic
properties of the particles, therefore, the breaking of symmetry induces
privileged directions in the space-time that can be measured experimentally,
and thus it can give us hints about a more fundamental theory behind the
broken symmetry. This proposal which at first was formulated to be a
renormalizable theory as cornerstone, it became known as Standard Model
Extension (SME) \cite{KostColl}.

Another possibility is to go beyond the SM by relaxing the
renormalizability, since we are searching for a more fundamental theory
through an effective theory, in a simplest way, by non minimal coupling which violates
power-counting renormalizability ($D_m=\partial _m+ieA_m+i\frac{g}{2}\epsilon _{mnab}\eta^nF^{ab}$), in an effective the Dirac equation. The proposal of such a non-minimal
coupling was suggested by \cite{nonmin} with a Lorentz-violating and CPT-odd
nonminimal coupling between fermions and the gauge field, which is carried
out by a Carrol-Field-Jackiw (CFJ) term \cite{CFJ}. This proposal has been
investigated in several distinct scenarios \cite%
{NM3,NMhall,NMmaluf,NMbakke,NMABC}.

Connections between high-energy theories and phenomena of condensed matter
proved to be very fertile and stimulated the development of new similar
mechanism in both areas. The BCS superconductivity is one of the remarkable
examples of such links. Nowadays graphene \cite{3,4}, which is described by
a 2+1 dimensional massless Dirac equation, is an interesting application for formulation of effective proposals of QED \cite{marino, van}. There is a
novel class of materials known as Weyl semi-metals \cite{20}, and in these
materials an effective low-energy theory is discussed in \cite{weyl} chosen 
to take the form of a Lorentz violating version of QED, which generates
the non minimal CPT-violating gauge coupling by the fermionic sector. This
interesting question that has been the target of research is the possibility
of perturbative generating the CFJ term has been addressed in
Refs. \cite{Radio1}, \cite{Radio2}, \cite{Radio3}.\ Also studies of
relativistic quantum effects \cite{knut1} that stem from a non-minimal
coupling with Lorentz symmetry breaking \cite{knut2} has opened the
possibility of investigating new implications in quantum mechanics that this
Lorentz-violating background can promote.

One of the main ideas along this line consists in a suggestion that
corresponding couplings in a purely gauge (or matter) sector could emerge as
a consequence of couplings of these fields with the spinor matter in some
underlying model. Proceeding in this way, the CFJ term has been successfully
generated on the base of a minimal Lorentz-breaking extension of the QED 
\cite{CFJ}. Further, the idea of using the nonminimal couplings as well has
been proposed, with the CPT-even aether-like term has been generated within
this prescription and showed to be finite on the base of the CPT-odd
coupling \cite{aether}. However, up to now, the question of use the CPT-even
couplings for the perturbative generation of the quantum correction is still
open. While such couplings have been proposed already in \cite{nonmin},
obtaining of quantum corrections on their base has not been studied enough.

Recently, an interesting example of the nonminimal CPT-even coupling has
been proposed in \cite{start} which was shown to be important and
interesting within the context of the scattering processes. Therefore, it is
natural to study the loop corrections generated by this coupling. Some
preliminary studies on this issue have been performed in \cite{start1}.
So, the natural continuation of the study consists, first, in
studying of higher-point functions, second, in studying of quantum
corrections to vertices of interaction. These problems will be addressed in
this paper.

\section{The CPT-even gauge theory}

Let us start with the following action of the extended QED involving the CPT-even term:
\bea
\label{action}
S=\int d^4x\bar{\psi} (i\ds-m-e\As+\lambda\kappa_{abcd}\sigma^{ab}F^{cd})\psi.
\eea
This action has been proposed in \cite{start}. It is evidently gauge invariant. Our signature is $(+---)$, and we use $\sigma^{ab}=\frac{i}{2}[\gamma^a,\gamma^b]$.

It is clear that this theory is unfortunately non-renormalizable as well as other theories with nonminimal gauge-matter coupling. In principle, to circumvent this difficulty, we can suggest the gauge field to be purely external, therefore, all quantum corrections are exhausted by the one-loop results, thus, adding appropriate nonlinear terms into the purely gauge sector we avoid all problems with the potential non-renormalizability. However, except of the subsection II.A where such an approach is sufficient to describe the relevant results, we suggest that we consider the effective theory to study the low-energy behaviour of some underlying fundamental theory, with the non-renormalizability arises for example as a consequence of integration over some extra fields. It is well known that in general, the interpretation of the non-renormalizable theory as an effective one for study of the infrared domain is very natural, the paradigmatic example is the four-fermion interaction. Different issues related to study of quantum aspects of non-renormalizable theories are discussed in papers \cite{nonren}. Moreover, since the Lorentz symmetry breaking naturally emerges in the low-energy limit of the string theory \cite{KS}, it is natural to expect that the resulting theory would be non-renormalizable just as occurs for example with the general chiral superfield model arising in the low-energy limit of superstring theory \cite{BCP}. 
Nevertheless, we can treat this theory as an effective one for the one-loop approximation.

\subsection{Effective action in a pure gauge sector}

In a pure gauge sector, the effective action in our theory is given by the following fermionic determinant:
\bea
\Gamma^{(1)}=i{\rm Tr}(i\ds-m-e\As+\lambda\kappa_{abcd}\sigma^{ab}F^{cd}).
\eea
Here we suggest the $\kappa_{abcd}$ to be dimensionless for simplicity, with the dimension of $\lambda$ is $-1$. Thus, the theory is non-renormalizable, so we treat it as an effective model for low-energy studies.
We start our consideration with a brief review of the results obtained earlier for this theory, that is, one-loop contributions to the two-point function of the gauge field which is represented by the Feynman diagram given by Fig. 1. 

\vspace*{2mm}

\begin{figure}[!h]
\begin{center}
\includegraphics[angle=0,scale=1.00]{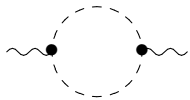}
\end{center}
\caption{Two-point function of the gauge field.}
\end{figure}

\vspace*{2mm}

There will be three different contributions to it: one with two external $A^a$ field, one mixed with one external $A^a$ and one external $F^{cd}$, and one with two external $F^{cd}$ fields.
As usual we have the Dirac propagator
\bea
<\psi(k)\bar{\psi}(-k)>=i(\ks-m)^{-1}.
\eea

The case where both of two external legs corresponds to the simple $A_a$ field is well-studied being Lorentz-invariant. It is the well-known two-point function of the gauge field in the usual QED (see f.e. \cite{Fuji}) which yields
\bea
\Sigma_{QED}=\frac{e^2}{24\pi^2\epsilon}F_{mn}F^{mn}.
\eea
The Lorentz-breaking corrections to the two-point function were found in \cite{start1}.  It was shown there that
the diagram with one external $A_a$ and one external $F_{cd}$ looks like, for $d=4+\epsilon$ (actually, our $\lambda$ is $1/2$ of $\lambda$ from \cite{start1}, and our $\epsilon$ is $\epsilon$ from \cite{start1}, multiplied by $-2$),
\bea
\label{i}
I=-\frac{1}{2\pi^2\epsilon}e\lambda m\kappa_{abcd}F^{ab}F^{cd}+{\rm fin}.
\eea
This is the lower-order aether-like contribution. Unfortunately, it diverges. In a particular case it reproduces the aether term found in \cite{aether}. Then, the diagram with two external $F_{cd}$ yields \cite{start1}
\bea
\label{j}
J=\frac{\lambda^2m^2}{2\pi^2\epsilon}(\kappa_{abcd}F^{cd})^2+{\rm fin}.
\eea
Again, this term is divergent. It is also aether-like. So, in the extended QED involving two couplings, the minimal and nonminimal ones, the aether-like terms naturally emerge both from the purely nonminimal and from the ``mixed'' sector. These terms are divergent. 
We consider the following explicit form of $\kappa_{abcd}$: 
\bea
\label{aethercase}
\kappa_{abcd}=k_{ac}\eta_{bd}-k_{ad}\eta_{bc}+k_{bd}\eta_{ac}-k_{bc}\eta_{ad},
\eea
For the first step, we consider $k_{ab}=u_au_b$
with $u_a$ is a dimensionless constant vector. The importance of this form of $\kappa_{abcd}$ consists, first, in its simplicity, second, in the fact that in this case one has the known CPT-even Lorentz-breaking term to be in the form $\kappa_{abcd}F^{ab}F^{cd}=4u^aF_{ab}u_cF^{cb}$, thus reproducing perfectly the aether term studied in \cite{Carroll,aether}, one finds that $\kappa_{abcd}\kappa^{cdmn}=2u^2\kappa_{ab}^{\phantom{ab}mn}$. 
 Let us choose the vector $u^a$ to be light-like, $u^2=u^au_a=0$, for the sake of simplicity, which annihilates the term of the second order in $\kappa_{abcd}$.
In this case, to achieve the one-loop renormalizability one should introduce the following free Lagrangian of the gauge field:
\bea
L_{gauge}=\Gamma_2=-\frac{1}{4}Z_1F_{ab}F^{ab}+\frac{1}{2}Z_2\kappa^{abcd}F_{ab}F_{cd},
\eea
with the renormalization constants are
\bea
Z_1=1+\frac{e^2}{6\pi^2\epsilon}, \quad\, 
Z_2=1-\frac{e\lambda m}{\pi^2\epsilon}.
\eea

We note that other choices for $k_{ab}$ can be discussed as well, for example the traceless one \cite{Cas2010}
\bea
\label{traceless}
k_{ab}=\frac{1}{2}(u_av_b+u_bv_a)-\frac{1}{4}g_{ab}(u\cdot v).
\eea
However, in this case one has f.e. $(\kappa_{abcd}F^{cd})^2=8F^{cn}F_{cd}(k^2)^d_n+8k_d^nk_c^mF^{cd}F_{mn}$, with $(k^2)_a^b=k_{an}k^{bn}$, and for the traceless $k_{ab}$, even for light-like $u_a$ and $v_a$, the $(\kappa_{abcd}F^{cd})^2$ will not vanish. In this case the sum of $I$ (\ref{i}) and $J$ (\ref{j}) will yield
\bea
I+J&=&-\frac{2}{\pi^2\epsilon}e\lambda m (4(F^{cd}u_d)(F_{cm}v^m)-F_{ab}F^{ab}(u\cdot v))+\\
&+&\frac{\lambda^2m^2}{\pi^2\epsilon}\left(u^2(F^{cd}v_d)^2+v^2(F^{cd}u_d)^2-2(F^{cd}u_cv_d)^2-2
(F^{cd}u_d)(F_{cm}v^m)(u\cdot v)\right)+{\rm fin}.\nonumber
\eea
We note again that for $k_{ab}=u_au_b$ with the light-like $u_a$ (i.e. $u_a=v_a$), the result will have the simplest form. Similarly, the same situation occurs as well for other contributions we will find.

There is one more divergence in four dimensions which has not ever considered. It is given by the four-point function. It is well known that the divergent part of the purely minimal contribution to the four-point function is zero, by the gauge invariance reasons. Actually, since the trace of the product of three $F_{mn}$ tensors is zero (and of three $\tilde{F}_{ab}=\kappa^{abcd}F_{cd}$ tensors as well, also by symmetry reasons, so, we note that for diagrams composed by purely nonminimal couplings there is a kind of Furry theorem, that is, the one-loop diagrams involving only $\tilde{F}$ legs and odd number of vertices identically vanishes), and forming of the $F_{mn}$ on the base of the external $A_m$ decreases the degree of divergence by one, we see that the only possible divergent contribution to this function consistent with the gauge invariance is purely non-minimal one, that is, involving namely four non-minimal vertices (with all gauge-invariant one-loop four-leg diagrams with at least one minimal vertex will be finite), and given by the expression
\bea
\label{ourtrace}
\Gamma_4&=&-\frac{1}{4}\int\frac{d^4k}{(2\pi)^4}\frac{1}{(k^2-m^2)^4}{\rm tr}[\sigma^{ab}(\ks+m)\sigma^{cd}(\ks+m)\sigma^{mn}(\ks+m)\sigma^{pq}(\ks+m)]
\times\nonumber\\&\times&
\tilde{F}_{ab}\tilde{F}_{cd}\tilde{F}_{mn}\tilde{F}_{pq}+{\rm fin}.
\eea
We can make permutations between external legs of this contribution, so we get
\bea
\Gamma_4&=&-\frac{1}{12}\int\frac{d^4k}{(2\pi)^4}\frac{1}{(k^2-m^2)^4}{\rm tr}[\sigma^{ab}(\ks+m)\sigma^{cd}(\ks+m)\sigma^{mn}(\ks+m)\sigma^{pq}(\ks+m)+\nonumber\\&+&
\sigma^{ab}(\ks+m)\sigma^{mn}(\ks+m)\sigma^{cd}(\ks+m)\sigma^{pq}(\ks+m)+\nonumber\\&+&
\sigma^{ab}(\ks+m)\sigma^{pq}(\ks+m)\sigma^{mn}(\ks+m)\sigma^{cd}(\ks+m)
]
\times\nonumber\\&\times&
\tilde{F}_{ab}\tilde{F}_{cd}\tilde{F}_{mn}\tilde{F}_{pq}+{\rm fin}.
\eea

Actually, to obtain the divergence only, we must calculate the leading term from $\Gamma_4$ which looks like
\bea
\Gamma_4&=&-\frac{1}{4}\tilde{F}_{ab}\tilde{F}_{cd}\tilde{F}_{mn}\tilde{F}_{pq}\int\frac{d^4k}{(2\pi)^4}
\frac{k_ek_fk_gk_h}{(k^2-m^2)^4}
\times\nonumber\\&\times&
{\rm tr}[\sigma^{ab}\gamma^e\sigma^{cd}\gamma^f\sigma^{mn}\gamma^g\sigma^{pq}\gamma^h]+{\rm fin}.
\eea
To proceed with the traces and to keep only the divergent contributions, we use the dimensional reduction approach, that is, we suggest that the Dirac matrices are defined in four dimensions, while the integrals are evaluated in $4+\epsilon$ dimensions. We use this prescription henceforth as we intend to fix the pole part only which is unambiguous in one-loop calculations. Then, it is well known that
\bea
\int\frac{d^{4+\epsilon}k}{(2\pi)^{4+\epsilon}}
\frac{k_ek_fk_gk_h}{(k^2-m^2)^4}=\frac{1}{3}\frac{1}{16\pi^2\epsilon}(\eta_{ef}\eta_{gh}+\eta_{eg}\eta_{fh}+\eta_{eh}\eta_{fg}),
\eea
and that $\gamma^m\sigma^{ab}\gamma_m=0$ in four dimensions. Taking into account all this, we find that only one term contributes to the trace:
\bea
\Gamma_4&=&-\frac{1}{192\pi^2\epsilon}\tilde{F}_{ab}\tilde{F}_{cd}\tilde{F}_{mn}\tilde{F}_{pq}
{\rm tr}[\sigma^{ab}\gamma^m\sigma^{cd}\gamma^n\sigma^{mn}\gamma_m\sigma^{pq}\gamma_n]+{\rm fin}.
\eea
It remains to find this trace. Straightforward calculation yields, if again $\epsilon=d-4$,
\bea
\label{gamma4}
\Gamma_4=-\frac{1}{4\pi^2\epsilon}(-\frac{2}{3}(\tilde{F}_{ab}\tilde{F}^{ab})^2+\frac{8}{3}\tilde{F}_{ab}\tilde{F}^{bc}\tilde{F}_{cd}\tilde{F}^{da})+{\rm fin},
\eea
which diverges as it must be. 

However, for the $k_{ab}$ described by the only light-like $u^a$, (\ref{gamma4})  vanishes. We see that for the light-like $u^a$, the only new divergence in a pure gauge sector is of the second order in $F_{ab}$, being aether-like.

For the $k_{ab}$ given by (\ref{traceless}), we get
\bea
\Gamma_4&=&-\frac{1}{4\pi^2\epsilon}\Big(-\frac{8}{3}\left[u^2(F^{cd}v_d)^2+v^2(F^{cd}u_d)^2-2(F^{cd}u_cv_d)^2-2
(F^{cd}u_d)(F_{cm}v^m)(u\cdot v)\right]^2+\nonumber\\ &+& \frac{8}{3}\tilde{F}_{ab}\tilde{F}^{bc}\tilde{F}_{cd}\tilde{F}^{da}\Big)+{\rm fin}.
\eea
We note that it is natural to suggest that the $k_{ab}$ is described by only light-like $u^a$, since the vanishing of the four-point divergent seems to be natural from the phenomenological viewpoint.

\subsection{The one-loop two-point function of the spinor field}

If we abandon the restriction that the gauge field is purely external, we should, as it has been suggested in \cite{start1}, consider the lower corrections to the two-point function of the spinor field and to the vertex function. 
One can easily verify that if we have only one non-minimal vertex with $\kappa_{abcd}$ given by (\ref{aethercase}), the one-derivative contribution to the two-point function of the spinor field will be purely Lorentz invariant and vanishing for the light-like $u^a$, being equal to $-\frac{e\lambda u^2}{4\pi^2\epsilon}\bar{\psi}(-p)(-3m \ps+2m^2)\psi(p)$,
So, let us consider the Feynman diagram formed by two nonminimal vertices. It is given by Fig. 2.

\vspace*{2mm}

\begin{figure}[!h]
\begin{center}
\includegraphics[angle=0,scale=1.00]{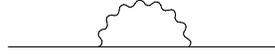}
\end{center}
\caption{Two-point function of the spinor field.}
\end{figure}

\vspace*{2mm}

We suggest that we use the Feynman gauge, so, the propagator of the gauge field is
\bea
<A^c(k)A^d(-k)>=i\frac{\eta^{cd}}{k^2}.
\eea
Then, one should take into account acting of the derivatives on the gauge fields in the vertices. The resulting contribution is
\bea
\Gamma_{\bar{\psi}\psi}=-\frac{\lambda^2}{2}\kappa_{abcd}\kappa_{a'b'c'd'}\int d^4x \bar{\psi}\sigma^{ab}<F^{cd}F^{c'd'}>\sigma^{a'b'}<\psi\bar{\psi}>\psi.
\eea
After the Fourier transform, we get
\bea
\Gamma_{\bar{\psi}\psi}&=&\frac{\lambda^2}{2}\kappa_{abcd}\kappa_{a'b'c'd'}\int \frac{d^dk}{(2\pi)^d}\frac{1}{[(k+p)^2-m^2]k^2} \bar{\psi}(-p)\sigma^{ab}(\ks+\ps+m)\sigma^{a'b'}\times\nonumber\\&\times&
\Big[k^ck^{c'}\eta^{dd'}+k^dk^{d'}\eta^{cc'}-k^ck^{d^{\prime}}\eta^{c^{\prime}d}
-k^dk^{c^{\prime}}\eta^{cd^{\prime}}\Big]
\psi(p).
\eea
It remains to find the integral. Keeping only the first order in the external momentum $p$ to get the generalization of the Dirac action, after simple changes of variables through Feynman representation, and taking into account the symmetries of the $\kappa$ coefficients, we find: 
\bea
\Gamma_{\bar{\psi}\psi}&=&\frac{\lambda^2}{2}\frac{4}{d}\kappa_{abcd}\kappa_{a'b'c'd'}\times\nonumber\\&\times&
\int_0^1dx \int \frac{d^dk}{(2\pi)^d}\frac{k^2}{(k^2-m^2x)^2}\Big[\bar{\psi}(-p)\sigma^{ab}[\ps(1-x)+m]\sigma^{a'b'}\psi(p)\eta^{cc'}\eta^{dd'}-
\nonumber\\&-&
\bar{\psi}(-p)\sigma^{ab}\gamma^m\sigma^{a'b'}\psi(p)(\delta^{c'}_mp^c+\delta^c_mp^{c'})\eta^{dd'}
\Big].
\eea
After integration over momenta, we have:
\bea
\Gamma_{\bar{\psi}\psi}&=&-\lambda^2\frac{(m^2)^{d/2-1}}{(4\pi)^{d/2}} \Gamma(1-\frac{d}{2})\kappa_{abcd}\kappa_{a'b'c'd'}\times\nonumber\\&\times&
\int_0^1dxx^{d/2-1}\Big[\bar{\psi}(-p)\sigma^{ab}[\ps(1-x)+m]\sigma^{a'b'}\psi(p)\eta^{cc'}\eta^{dd'}-
\nonumber\\&-&
\bar{\psi}(-p)\sigma^{ab}\gamma^m\sigma^{a'b'}\psi(p)(\delta^{c'}_mp^c+\delta^c_mp^{c'})\eta^{dd'}
\Big].
\eea
This expression yields at $d=4+\epsilon$:
\bea
\Gamma_{\bar{\psi}\psi}&=&-\frac{\lambda^2m^2}{16\pi^2\epsilon}\kappa_{abcd}\kappa_{a'b'c'}^{\phantom{a'b'c'}d}
\Big[\bar{\psi}(-p)\sigma^{ab}[\frac{1}{3}\ps+m]\sigma^{a'b'}\psi(p)\eta^{cc'}-
\nonumber\\&-&
\bar{\psi}(-p)\sigma^{ab}\gamma^m\sigma^{a'b'}\psi(p)(\delta^{c'}_mp^c+\delta^c_mp^{c'})
\Big]+{\rm fin}.
\eea
For $\kappa_{abcd}$ given by (\ref{aethercase}), we find
\bea
\Gamma_{\bar{\psi}\psi}&=&\frac{\lambda^2m^2}{16\pi^2\epsilon}
\Big[8u^2u_au_{a'}\bar{\psi}(-p)\sigma^{ab}[\frac{1}{3}\ps+m]\sigma^{a'}_{\phantom{a'}b}\psi(p)-
\\&-&
4u_au_{a'}u_cu_{c'}
\bar{\psi}(-p)\sigma^{ab}\gamma^m\sigma^{a'}_{\phantom{a'}b}\psi(p)(\delta^{c'}_mp^c+\delta^c_mp^{c'})-\nonumber\\
&-&
4u^2u_au_{a'}\bar{\psi}(-p)
\sigma^a_{\phantom{a}c}\gamma^m\sigma^{a'}_{\phantom{a'}c'}\psi(p)(\delta^{c'}_mp^c+\delta^c_mp^{c'})
\Big]+{\rm fin}.\nonumber
\eea
It can be simplified even more. We use the fact that
\bea
\sigma^{ab}\gamma^m\sigma^{a'}_{\phantom{a'}b}=-2\gamma^a\eta^{ma'}-2\gamma^{a'}\eta^{ma}+\gamma^m\gamma^a\gamma^{a'}.
\eea
As a result we get
\bea
\label{2pfin}
\Gamma_{\bar{\psi}\psi}&=&\frac{\lambda^2m^2u^2}{6\pi^2\epsilon}
\bar{\psi}(-p)\Big[25u^2\ps+9u^2m-10\us (u\cdot p)
\Big]\psi(p)+{\rm fin},
\eea
which vanishes for the light-like $u^a$. So, there is no one-loop contribution to the two-point function of the spinor field in our case.

Again, if the $k_{ab}$ is given by (\ref{traceless}), we get
\bea
\Gamma_{\bar{\psi}\psi}&=&-\frac{\lambda^2m^2}{16\pi^2\epsilon}
\Big[\bar{\psi}(-p)\sigma^{ab}[\frac{1}{3}\ps+m]\sigma^{a'b'}\psi(p)\times\nonumber\\&\times&
[2k_{ac}k_{a'}^c\eta_{bb'}+2k_{bc}k_{b'}^c\eta_{aa'}-2k_{ac}k_{b'}^c\eta_{ba'}-2k_{bc}k_{a'}^c\eta_{ab'}+4k_{aa'}k_{bb'}-4k_{ab'}k_{ba'}]-
\nonumber\\ &-&
\kappa_{abcd}\kappa_{a'b'c'}^{\phantom{a'b'c'}d}\bar{\psi}(-p)\sigma^{ab}\gamma^m\sigma^{a'b'}\psi(p)(\delta^{c'}_mp^c+\delta^c_mp^{c'})
\Big]+{\rm fin}.
\eea
Taking into account the antisymmetry of $\sigma^{ab}$ with respect to indices $a$ and $b$, we get
\bea
\Gamma_{\bar{\psi}\psi}&=&-\frac{\lambda^2m^2}{16\pi^2\epsilon}
\Big[8\bar{\psi}(-p)\sigma^{ab}[\frac{1}{3}\ps+m]\sigma^{a'b'}\psi(p)
[2k_{ac}k_{a'}^c\eta_{bb'}+k_{aa'}k_{bb'}]-
\nonumber\\ &-&
\kappa_{abcd}\kappa_{a'b'c'}^{\phantom{a'b'c'}d}\bar{\psi}(-p)\sigma^{ab}\gamma^m\sigma^{a'b'}\psi(p)(\delta^{c'}_mp^c+\delta^c_mp^{c'})
\Big]+{\rm fin},
\eea
or, as is the same,
\bea
\Gamma_{\bar{\psi}\psi}&=&-\frac{\lambda^2m^2}{16\pi^2\epsilon}
\Big[8\bar{\psi}(-p)\sigma^{ab}[\frac{1}{3}\ps+m]\sigma^{a'b'}\psi(p)
[2k_{ac}k_{a'}^c\eta_{bb'}+k_{aa'}k_{bb'}]-
\nonumber\\ &-&
4(k_{ac}k_{a'c'}\eta_{bb'}-k_{ab'}k_{a'c'}\eta_{bc}-k_{ac}k_{a'b}\eta_{b'c'}+k_{ad}k_{a'}^d\eta_{bc}\eta_{b'c'})
\times\nonumber\\&\times&
\bar{\psi}(-p)\sigma^{ab}
(\gamma^{c'}p^c+\gamma^cp^{c'})
\sigma^{a'b'}\psi(p)
\Big]+{\rm fin}.
\eea
This is the final result for the two-point function of $\psi$. We note that the requirement for $k_{ab}$ to be traceless does not simplify the result essentially.

\subsection{The one-loop three-point vertex function}

Now, we can find the three-point vertex function in the theory which evidently displays the quadratic divergence. It is given by the Feynman diagram:

\begin{figure}[!h]
\begin{center}
\includegraphics[angle=0,scale=1.00]{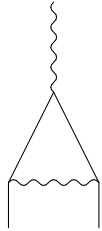}
\end{center}
\caption{Three-point function.}
\end{figure}

Following the prescriptions of the derivative expansion, it is natural to restrict ourselves only by the contributions of first order in derivatives, like our triple vertex 
$\lambda\kappa_{abcd}\bar{\psi}\sigma^{ab}F^{cd}\psi$. 

We start with considering the Feynman diagram of first order in $\lambda$, we will have two situations.
First, let the nonminimal vertex is associated with external gauge leg. In this case, it yields
\bea
\Gamma_1=\frac{\lambda e^2}{2}\int\frac{d^4k}{(2\pi)^4}\bar{\psi}\frac{\gamma^m(\ks+m)\sigma^{ab}(\ks+m)\gamma^n}{(k^2-m^2)^2}\psi\frac{\eta_{mn}}{k^2}\tilde{F}_{ab},
\eea 
where the external momenta are suppressed being irrelevant within our approximation.
By symmetry reasons, there are two nontrivial contributions in the numerator: one of zero order in momenta, 
$m^2\gamma^m\sigma^{ab}\gamma_m$, which vanishes since $\gamma^m\sigma^{ab}\gamma_m=0$ in four dimensions, and another of second order in momenta, $\gamma^m\ks\sigma^{ab}\ks\gamma_m$. However, using the definition of $\ks$, one can easily rewrite this factor as $k_pk_q\gamma^m\gamma^p\sigma^{ab}\gamma^q\gamma_m$, and under the integral one can replace $k_pk_q\to\frac{1}{4}\eta_{pq}k^2$, which allows to write our factor as $\frac{k^2}{4}\gamma^m\gamma^p\sigma^{ab}\gamma_p\gamma_m=0$. So, in this case the three-point function yields a zero result. 

Now we consider the case when the only nonminimal vertex is associated to the external spinor leg. The corresponding contribution is, for $p=-p_1-p_2$, with two terms correspond to two choices of a position of the nonminimal vertex:
\bea
\label{gamma3}
\Gamma_3^{(1)}&=&\frac{1}{2}\lambda e^2\kappa^{nlab}\int\frac{d^4k}{(2\pi)^4}\frac{(k_l-p_{1l})}{(k^2-m^2)(k-p_1)^2[(k+p)^2-m^2]}
\times\nonumber\\ &\times&
\bar{\psi}(p_1)\Big[\gamma^m(\ks+m)\gamma^r(\ks+\ps+m)\sigma_{ab}- \nonumber\\&-&
\sigma_{ab}(\ks+m)\gamma^r(\ks+\ps+m)\gamma^m
\Big]\psi(p_2)\eta_{mn}A_r(p)+(p_1\leftrightarrow p_2).
\eea
Here we introduced two terms with interchange between $p_1$ and $p_2$ since, after the substitution into the effective action, the integration over both these momenta should be done.
Actually, we want to find only the divergence of this expression. It is clear thus that the terms proportional to $m^2$ in the numerator will be irrelevant due to their finiteness. Also, the desired expression will involve the even number of Dirac matrices because of structure of contractions (actually, we look for the structure like $\bar{\psi}\sigma^{ma}\psi F_{ab}u_mu^b$, that is, the lower Lorentz-breaking gauge invariant contribution involving two $u^m$ vectors). Thus, (\ref{gamma3}) is reduced to
\bea
\label{gamma3a}
\Gamma_3^{(1)}&=&\frac{\lambda e^2}{2}\kappa^{nlab}\int\frac{d^4k}{(2\pi)^4}\frac{(k_l-p_{1l})}{(k^2-m^2)(k-p_1)^2[(k+p)^2-m^2]}
\times\nonumber\\ &\times&
\bar{\psi}(p_1)\Big[\gamma_n\ks\gamma^r(\ks+\ps)\sigma_{ab}- 
\sigma_{ab}\ks\gamma^r(\ks+\ps)\gamma_n
\Big]\psi(p_2)A_r(p)+(p_1\leftrightarrow p_2).
\eea
The straightforward calculations show that the relevant result is
\bea
\label{e28}
\Gamma_3^{(1)}&=&\frac{1}{4\pi^2\epsilon}\lambda e^2\bar{\psi}\sigma^{ma}\psi F_{ab}u_mu^b+\frac{e^2\lambda m}{2\pi^2\epsilon}
\bar{\psi}[(A\cdot u)\us+\frac{1}{2}\As u^2]\psi+{\rm fin}.
\eea
Again, if the $k_{ab}$ is given by (\ref{traceless}), we get
\bea
\Gamma_3^{(1)}&=&\frac{1}{8\pi^2\epsilon}\lambda e^2\bar{\psi}\sigma^{ma}F_{ab}(u_mv^b+v_mu^b)\psi+\frac{e^2\lambda m}{4\pi^2\epsilon}
\bar{\psi}[(A\cdot u)\vs+(A\cdot v)\us]\psi+{\rm fin}.
\eea
Treating the one-loop corrections to the three-point function with more insertions of nonminimal vertices, we easily find that they will be either finite involving larger numbers of derivatives applied to external gauge leg, or involving the contractions like $(u^2)^n$ which vanish for the light-like $u^m$. We note that the first contribution to this term, being proportional to $\bar{\psi}\sigma^{ma}\psi F_{ab}u_mu^b$, perfectly reproduces the structure of the nonminimal vertex of our action, $\kappa^{abcd}\bar{\psi}\sigma^{ab}\psi F_{cd}$, for the $\kappa_{abcd}$ given by (\ref{aethercase}), whereas the second one involves an  $A_m$-dependent part of the aether-extended Dirac Lagrangian \cite{Carroll} $i\bar{\psi}u^mu^n\gamma_mD_n\psi$.

The case when two vertices are nonminimal is given by two graphs. In one of them, the external gauge leg is attached to the minimal vertex.
The contribution of this graph is
\bea
\Gamma_{3a}^{(2)}&=&\frac{e\lambda^2}{2}\kappa_{aba'b'}\kappa_{efe'f'}\int\frac{d^dp_1d^dp_2}{(2\pi)^{2d}}A_m(-p_1-p_2)
\times\nonumber\\&\times&
\int\frac{d^dk}{(2\pi)^d}\bar{\psi}(p_1)\sigma^{a'b'}(\ks+m)\gamma^m (\ks-\ps_1-\ps_2+m)\sigma^{e'f'}\psi(p_2)\times\nonumber\\&\times&
\frac{1}{(k^2-m^2)[(k-p_1-p_2)^2-m^2](k-p_1)^2}
\times\nonumber\\&\times&
[(k-p_1)^a(k-p_1)^e\eta^{bf}+(k-p_1)^b(k-p_1)^f\eta^{ae}-\nonumber\\&-&
(k-p_1)^a(k-p_1)^f\eta^{be}-(k-p_1)^b(k-p_1)^e\eta^{af}]+(p_1\leftrightarrow p_2).
\eea
The exact result is very cumbersome, so we give only its form at $u^2=0$ it yields
\bea
\Gamma_{3a}^{(2)}&=&\frac{e\lambda^2}{3\pi^2\epsilon}\Big[
(u\cdot\pa)^2\bar{\psi}u^aA_a\us\psi +
\bar{\psi}u^aA_a\us(u\cdot\pa)^2\psi-
(u\cdot\pa)\psi u^aA_a(u\cdot\pa)\us\psi
\Big].
\eea
Again, if the $k_{ab}$ is given by (\ref{traceless}), we get
\bea
\Gamma_{3a}^{(2)}&=&\frac{e\lambda^2m}{\pi^2\epsilon}\Big[(u\cdot\pa)\bar{\psi}v^aA_a\us\vs\psi - (v\cdot\pa)\bar{\psi}u^aA_a\us\vs\psi + \bar{\psi}v^aA_a\us\vs(u\cdot\pa)\psi - \bar{\psi}u^aA_a\us\vs(v\cdot\pa)\psi\Big] \nonumber\\
&& + \frac{e\lambda^2}{6\pi^2\epsilon}\Big[
(u\cdot\pa)(v\cdot\pa)\bar{\psi}u^aA_a\vs\psi + (v\cdot\pa)(u\cdot\pa)\bar{\psi}v^aA_a\us\psi +
\bar{\psi}u^aA_a\vs(u\cdot\pa)(v\cdot\pa)\psi \nonumber\\
&& + \bar{\psi}v^aA_a\us(v\cdot\pa)(u\cdot\pa)\psi - (u\cdot\pa)\psi v^aA_a(u\cdot\pa)\vs\psi - (v\cdot\pa)\psi u^aA_a(v\cdot\pa)\us\psi
\Big].
\eea
However, this result is of the second order in derivatives  thus being irrelevant in our approximation.

In another case, the external gauge leg is nonminimal:
\bea
\label{gamma3b}
\Gamma_{3b}^{(2)}&=&\frac{1}{2}e\lambda^2
\kappa^{nlab}\int\frac{d^4k}{(2\pi)^4}\frac{(k_l-p_{1l})}{(k^2-m^2)(k-p_1)^2[(k+p)^2-m^2]}
\times\nonumber\\ &\times&
\bar{\psi}(p_1)\Big[\gamma^m(\ks+m)\sigma^{rs}(\ks+\ps+m)\sigma_{ab}- \nonumber\\&-&
\sigma_{ab}(\ks+m)\sigma^{rs}(\ks+\ps+m)\gamma^m
\Big]\psi(p_2)\eta_{mn}\tilde{F}_{rs}(p)+(p_1\leftrightarrow p_2).
\eea
Since in this case it is sufficient to disregard momenta $p_1,p_2$ in internal lines, we have
\bea
\Gamma_{3b}^{(2)}&=&\frac{1}{2}e\lambda^2\kappa^{nlab}\int\frac{d^dk}{(2\pi)^d}\frac{k_l}{(k^2-m^2)^2k^2}
\times\nonumber\\ &\times&
\bar{\psi}(p_1)\Big[\gamma^m(\ks-m)\sigma^{rs}(\ks-m)\sigma_{ab}- 
\sigma_{ab}(\ks-m)\sigma^{rs}(\ks-m)\gamma^m
\Big]\psi(p_2)\eta_{mn}\tilde{F}_{rs}(p)+\nonumber\\&+&(p_1\leftrightarrow p_2).
\eea
The result is
\bea
\Gamma_{3b}^{(2)}&=&-\frac{e\lambda^2}{2\pi^2\epsilon}\Big[\pa^b\bar{\psi}u^a\tilde{F}_{ab}\us{\psi}+\bar{\psi}u^a\tilde{F}_{ab}\us\pa^b\psi+2m\bar{\psi}u^a\tilde{F}_{ra}\us\gamma^r\psi
\Big].
\eea
Also, we can integrate by parts to obtain
\bea
\Gamma_{3b}^{(2)}&=&\frac{e\lambda^2}{2\pi^2\epsilon}\Big[\bar{\psi}\us{\psi}u^a\pa^b\tilde{F}_{ab}-2m\bar{\psi}u^a\tilde{F}_{ra}\us\gamma^r\psi
\Big].
\eea
Again, if the $k_{ab}$ is given by (\ref{traceless}), we get
\bea
\Gamma_{3b}^{(2)}&=&\frac{e\lambda^2}{4\pi^2\epsilon}\Big[\bar{\psi}\us{\psi}u^a\pa^b\tilde{F}_{ab}-2m\bar{\psi}u^a\tilde{F}_{ra}\us\gamma^r\psi+\bar{\psi}\slashed{v}{\psi}v^a\pa^b\tilde{F}_{ab}-2m\bar{\psi}v^a\tilde{F}_{ra}\slashed{v}\gamma^r\psi
\Big].
\eea

We note that $u^a\tilde{F}_{ab}=0$ for the light-like $u_a$, thus, this expression completely vanishes (moreover, its first term is irrelevant since does not contribute to the first order of the derivative expansion). 

Also, as an illustration, let us consider the case when all three vertices are nonminimal.
The contribution of this graph is
\bea
\Gamma_3^{(3)}&=&\frac{\lambda^3}{2}\kappa_{aba'b'}\kappa_{efe'f'}\int\frac{d^dp_1d^dp_2}{(2\pi)^{2d}}\tilde{F}_{cd}(-p_1-p_2)
\times\nonumber\\&\times&
\int\frac{d^dk}{(2\pi)^d}\bar{\psi}(p_1)\sigma^{a'b'}(\ks+m)\sigma^{cd}(\ks-\ps_1-\ps_2+m)\sigma^{e'f'}\psi(p_2)\times\nonumber\\&\times&
\frac{1}{(k^2-m^2)[(k-p_1-p_2)^2-m^2](k-p_1)^2}
\times\nonumber\\&\times&
[(k-p_1)^a(k-p_1)^e\eta^{bf}+(k-p_1)^b(k-p_1)^f\eta^{ae}-\nonumber\\&-&
(k-p_1)^a(k-p_1)^f\eta^{be}-(k-p_1)^b(k-p_1)^e\eta^{af}].
\eea
Since our aim consists in obtaining only of a (divergent) correction to the vertex, with no more derivatives acting on any of the external fields, we can put $p_1,p_2\simeq 0$ within this integral, and do the inverse Fourier transform. We get
\bea
\Gamma_3^{(3)}&=&\frac{\lambda^3}{2}\kappa_{aba'b'}\kappa_{efe'f'}\int d^dx\tilde{F}_{cd}(x)
\times\nonumber\\&\times&
\int\frac{d^dk}{(2\pi)^d}\bar{\psi}(x)\sigma^{a'b'}(\ks+m)\sigma^{cd}(\ks+m)\sigma^{e'f'}\psi(x)\times\nonumber\\&\times&
\frac{1}{(k^2-m^2)^2k^2}
[k^ak^e\eta^{bf}+k^bk^f\eta^{ae}- 
k^ak^f\eta^{be}-k^bk^e\eta^{af}].
\eea
We can rewrite this term as a sum:
\bea
\Gamma_3^{(3)}=\Gamma_{31}+\Gamma_{32},
\eea
where
\bea
\Gamma_{31}&=&\frac{\lambda^3}{2}\kappa_{aba'b'}\kappa_{efe'f'}\int d^dx\tilde{F}_{cd}(x)
\times\nonumber\\&\times&
\bar{\psi}(x)\sigma^{a'b'}\gamma^m\sigma^{cd}\gamma^n\sigma^{e'f'}\psi(x)\times\nonumber\\&\times&
\int\frac{d^dk}{(2\pi)^d}\frac{k_mk_n}{(k^2-m^2)^2k^2}
[k^ak^e\eta^{bf}+k^bk^f\eta^{ae}- 
k^ak^f\eta^{be}-k^bk^e\eta^{af}];\nonumber\\
\Gamma_{32}&=&\frac{m^2\lambda^3}{2}\kappa_{aba'b'}\kappa_{efe'f'}\int d^dx\tilde{F}_{cd}(x)
\times\nonumber\\&\times&
\bar{\psi}(x)\sigma^{a'b'}\sigma^{cd}\sigma^{e'f'}\psi(x)\times\nonumber\\&\times&
\int\frac{d^dk}{(2\pi)^d}\frac{1}{(k^2-m^2)^2k^2}
[k^ak^e\eta^{bf}+k^bk^f\eta^{ae}- 
k^ak^f\eta^{be}-k^bk^e\eta^{af}].
\eea
To find the integral, we carry out the following replacement in $d$ dimensions:
\bea
\label{metrid}
k_ak_b&=&\frac{1}{d}\eta_{ab}k^2\nonumber\\
k_ak_bk_ck_d&=&\frac{1}{3d^2}(\eta_{ab}\eta_{cd}+\eta_{ac}\eta_{bd}+\eta_{ad}\eta_{bc})k^4.
\eea
As a result, we get
\bea
\Gamma_{31}&=&\frac{\lambda^3}{6d^2}\kappa_{aba'b'}\kappa_{efe'f'}\int d^dx\tilde{F}_{cd}(x)
\times\nonumber\\&\times&
\bar{\psi}(x)\sigma^{a'b'}\gamma_{m}\sigma^{cd}\gamma_{n}\sigma^{e'f'}\psi(x)
\int\frac{d^dk}{(2\pi)^d}\frac{k^2}{(k^2-m^2)^2}\times\nonumber\\&\times&
[\eta^{bf}(\eta^{mn}\eta^{ae}+\eta^{m a}\eta^{n e}+\eta^{m e}\eta^{n a})+
\eta^{ae}(\eta^{mn}\eta^{bf}+\eta^{\mu b}\eta^{n f}+\eta^{m f}\eta^{n b})-\nonumber\\
&-&\eta^{af}(\eta^{mn}\eta^{be}+\eta^{m b}\eta^{n e}+\eta^{m e}\eta^{n b})-
\eta^{be}(\eta^{mn}\eta^{af}+\eta^{m a}\eta^{n f}+\eta^{m f}\eta^{n a})
]
;\nonumber\\
\Gamma_{32}&=&\frac{m^2\lambda^3}{d}\kappa_{aba'b'}\kappa_{efe'f'}\int d^dx\tilde{F}_{cd}(x)\bar{\psi}(x)\sigma^{a'b'}\sigma^{cd}\sigma^{e'f'}\psi(x)
\times\nonumber\\&\times&
\int\frac{d^dk}{(2\pi)^d} 
\frac{1}{(k^2-m^2)^2}
[\eta^{ae}\eta^{bf}-\eta^{af}\eta^{be}].
\eea
We can integrate, which yields in $d=4+\epsilon$ dimensions, with taking into account the fact that $\gamma^{m}\sigma^{ab}\gamma_{m}=0$, second, the symmetric properties of the $\kappa_{abef}$:
\bea
\Gamma_{31}&=&-\frac{m^2\lambda^3}{384\pi^2\epsilon}\kappa_{aba'b'}\kappa_{efe'f'}\int d^4x\tilde{F}_{cd}(x)
\times\nonumber\\&\times&
\bar{\psi}(x)\sigma^{a'b'}\gamma_{m}\sigma^{cd}\gamma_{n}\sigma^{e'f'}\psi(x).
\eta^{bf}[\eta^{m a}\eta^{n e}+\eta^{m e}\eta^{n a}]
;\nonumber\\
\Gamma_{32}&=&-\frac{m^2\lambda^3}{64\pi^2\epsilon}\kappa_{aba'b'}\kappa_{efe'f'}\int d^4x\tilde{F}_{cd}(x)\bar{\psi}(x)\sigma^{a'b'}\sigma^{cd}\sigma^{e'f'}\psi(x)
\times\nonumber\\&\times&
[\eta^{ae}\eta^{bf}-\eta^{af}\eta^{be}].
\eea
Simplifying these expressions even more for $\kappa_{abcd}$ given by (\ref{aethercase}), we get
\bea
\Gamma_{31}&=&-\frac{m^2\lambda^3u^2}{48\pi^2\epsilon}\int d^4x\bar{\psi}(x)(-10u^2\sigma^{cd}-12(\sigma^{dm}u^cu_m-\sigma^{cm}u^du_m))\tilde{F}_{cd}(x)\psi(x);\nonumber\\
\Gamma_{32}&=&\frac{m^2\lambda^3u^4}{64\pi^2\epsilon}\int d^4x\bar{\psi}(x)\sigma^{cd}\tilde{F}_{cd}(x)\psi(x).
\eea
Thus, we found the result or the triple vertex explicitly. It vanishes for the light-like $u_a$. Moreover, the essentially Lorentz-breaking contribution to this vertex is
\bea
\Gamma^{(3)}_{3,LV}&=&\frac{m^2\lambda^3u^2}{4\pi^2\epsilon}\int d^4x\bar{\psi}(x)(\sigma^{dm}u^cu_m-\sigma^{cm}u^du_m)\tilde{F}_{cd}(x)\psi(x)
\eea
We see that this expression, first, does not involve derivatives of $F_{mn}$, second, vanishes at $u^2=0$. Again, if the $k_{ab}$ is given by (\ref{traceless}), we get
\bea
\Gamma^{(3)}_{3,LV}&=&\frac{2m^2\lambda^3}{\pi^2\epsilon}\int d^4x\bar{\psi}(x)\us\vs u_a v_b\tilde{F}^{ab}(x)\psi(x),
\eea
where we have not considered higher derivative terms.

We conclude that for the light-like $u^a$, the only relevant (that is, involving only first order in derivatives) result for the three-point function is given by (\ref{e28}) which in the massless limit yields
\bea
\label{e28a}
\Gamma_3&=&\frac{1}{4\pi^2\epsilon}\lambda e^2\bar{\psi}\sigma^{ma}\psi F_{ab}u_mu^b+{\rm fin}.
\eea
So, in this way we have been carried out the calculations of an
effective nonminimally coupling Dirac equation and, by radiative corrections
we have generated a new way to obtain the CPT-even term of the SME gauge
sector. The expressions of lowest order graphs (self-energy and the
three vertex-like contributions) have been obtained. As in four-dimensional
space-time the theory presents divergent terms we infer that the model
presents a low-energy domain.

\section{Problem of renormalization}

Let us discuss the problem of renormalization of our theory. Clearly, the model described by the action (\ref{action}) is non-renormalizable since the coupling $\lambda$ has a mass dimension equal to $-1$. The superficial degree of divergence of our theory is
\bea
\omega=4-E_A-\frac{3}{2}E_{\psi}+V_{\lambda}-N_d,
\eea
where $V_{\lambda}$ is a number of vertices proportional to $\lambda$, $E_A$ and $E_{\psi}$ are numbers of external $A$ and $\psi$ fields, and $N_d$ is a number of derivatives associated to external legs. Therefore, in principle, the superficial degree of divergence increases as the complexity of Feynman diagrams grows.

One of the manners to deal with this theory consists in restricting the study to the fermion determinant. In this case, the derivatives in the nonminimal vertices cannot act to the propagators, so, the maximal divergence is of the $F^4$ type. However, this approach restricts the theory very strongly. 

It is interesting to discuss the renormalizability of our theory. Although the theory is formally non-renormalizable, we note the following facts. 

First, if we consider the $\kappa_{abcd}$ of the form given by Eq. (\ref{aethercase}) with the Lorentz-breaking vector $u^a$ be light-like, many of quantum corrections simply vanish as occurs in all cases considered by us with corrections of second and higher orders in $\kappa_{abcd}$ (actually it means that only one vector $u^a$ can be contracted with a Dirac matrix, otherwise, using the commutation relations, one can obtain $u^au^b\gamma_a\gamma_b=u^2$ which vanishes for the light-like $u_a$; within this paper we observed for many one-loop examples that if a Feynman diagram involves at least two nonminimal vertices, its contribution will be proportional to $u^2$ and thus vanishes for the light-like $u_a$, and it is natural to expect that the similar situation will occur in higher loops).  Second, if we have two or more insertions of $\kappa_{abcd}$, in principle, after integration by parts, in principle we can arrive at terms like $\bar{\psi}\sigma^{ab}\psi(u\cdot\pa)^{2n}F_{bc}u_au^c$, but if we suggest that the background fields satisfy the condition $\pa_aF_{bc}=0$ (i.e. constant electric or magnetic fields), these results vanish. As a result, we conclude that under our restrictions many divergent contributions simply disappear.

\section{Conclusion}

In this paper, we succeeded to generate the SME sector gauge starting from
an effective fermionic theory with non-minimal coupling. While, earlier \cite{start1} the CPT-even term of the SME electrodynamics $(K_{F})_{mnpq}F^{mn}F^{pq}$ was generated in this model, now
we have calculated explicitly the lower corrections to the two-point
function of the spinor field and the three-point vertex function.

It turns out to be that, as a result, the aether-like terms
naturally arise not only in the pure gauge sector, but also in the matter sector, with the lower aether-like contribution in this sector is generated by the diagram with two nonminimal vertices, but it vanishes for the special choice of $\kappa_{abcd}$ of the form (\ref{aethercase}), with the tensor $k_{ab}$ completely described by a light-like $u^a$. 
Unfortunately, in four-dimensional space-time all these results are
divergent unlike those ones obtained on the base of the CPT-odd coupling 
\cite{aether}. However, the presence of these divergences, from one side,
can signalize the presence of the corresponding terms in a complete action
of the electromagnetic field, or, if we treat the Lorentz-breaking theory as
an effective field theory aimed for study of the low-energy domain, instead
of the divergences we can have the aether-like contributions depending on
the cutoff. 

Also, one can notice that the higher-point functions of the gauge field can
give origin to the effect of the photon splitting \cite{split}, or to the
generalized Heisenberg-Euler effective action of the nonlinear
electrodynamics. These higher-point functions, as well as other nonminimal Lorentz-breaking terms, can be also experimentally estimated with use of methods discussed in \cite{datatables}.

We note that the choice of the $\kappa_{abcd}$ of the form (\ref{aethercase}) described by light-like $u^a$ is crucial, it allows to simplify strongly the results, annihilating many contributions, which does not occur in other cases, for example, if two independent Lorentz-breaking vectors $u^a$ and $v^a$ are introduced. Apparently, it means that in this case no more than one nonminimal vertex is allowed for the Feynman diagram with a divergent contribution, and if this vertex is associated to the external gauge leg, the corresponding Feynman diagram can only yield the contribution of the form similar to the (\ref{e28a}), contributing only to renormalization of our nonminimal vertex. It is natural to expect that the situation in higher loops will be similar. We will consider higher loops and the case $u^2\neq 0$ in a next paper.

{\bf Acknowledgements.} Authors are grateful to R. Casana, M. M. Ferreira and F. E. P. dos Santos for important discussions. This work was partially supported by Conselho
Nacional de Desenvolvimento Cient\'{\i}fico e Tecnol\'{o}gico (CNPq). The work by A. Yu. P. has been supported by the
CNPq project No. 303783/2015-0.

\end{document}